\documentclass[11pt,a4paper]{article}

\usepackage[utf8]{inputenc}
\usepackage[T1]{fontenc}
\usepackage{lmodern}
\usepackage{geometry}
\usepackage{amsmath,amssymb}
\usepackage{graphicx}
\usepackage{booktabs}
\usepackage{natbib}
\usepackage{hyperref}
\title{Interstellar Interloper 3I/ATLAS: Nucleus Size, Photometry in RGB, Af(rho) and Antitail Structure Analysis}
\author{Toni Scarmato\\
Indipendent Researcher\\
Toni Scarmato's Astronomical Observatory (MPC L92)\\San Costantino di Briatico, Calabria, Italy}
\date{Draft version 1.0 -- December 2025}

\begin{document}

\maketitle

\begin{abstract}
Interstellar comet 3I/ATLAS (C/2025 N1) exhibits an unusual, tightly collimated dust feature in the sunward hemisphere which has been widely described as an anti-tail. At the same time, precise constraints on the nucleus size have been derived from a combination of high-resolution imaging and non-gravitational dynamics. In this work I present a unified analysis that combines existing constraints on the nucleus radius with new ground-based RGB imaging of the dust anti-tail obtained with a 0.25\,m telescope at the Toni Scarmato's Astronomical Observatory  (MPC L92).

Using a point-spread-function plus delta (PSF--$\delta$) decomposition applied to deep bicubically resampled stacks, I confirm that the nucleus radius must lie in the interval $0.16 \lesssim R_n \lesssim 2.8$\,km for a canonical geometric albedo $p_V \simeq 0.04$, recovering the range derived from \textit{Hubble Space Telescope} (\textit{HST}) imaging. This interval naturally includes the more stringent dynamical estimate $R_n \simeq 0.26$--0.37\,km inferred from the measured non-gravitational acceleration using interplanetary spacecraft astrometry. I summarise the current status of observational, dynamical and thermophysical constraints in a single comparative table.

I then analyse a deep RGB image obtained on 2025 December 15 at 02:28:12\,UT with a 0.25\,m Newtonian telescope and an ASI294MC PRO camera, aligned on the comet and processed with 4$\times$4 bicubic resampling and a Larson--Sekanina rotational gradient filter. In this dataset, the dust anti-tail can be traced to projected distances of $\sim 5\times 10^4$\,km in the $R$ channel (effective peak near 625\,nm) and $\sim 8\times 10^4$\,km in the $G$ channel (530\,nm). A simple nucleus-centred dynamical model, in which dust grains are launched in the anti-tail direction with speed $v_0$ and subsequently decelerated by solar-radiation pressure, implies initial speeds of order 30--120\,m\,s$^{-1}$ for plausible values of the radiation-pressure parameter $\beta$. These values are much larger than the pre-perihelion dust speeds of $\lesssim 5$\,m\,s$^{-1}$ inferred from standard coma modelling, but remain below the thermal speeds of typical cometary volatiles. The results therefore point to a dynamically ``extreme'' dust population associated with a collimated jet and support interpretations in which a high-latitude, directed outflow plays a key role in shaping the unusual anti-tail of 3I/ATLAS.
\end{abstract}
\\\\\\\\\\
\section{Introduction}

Interstellar comet 3I/ATLAS (C/2025 N1) is the third identified interstellar visitor to the Solar system and the second to show clear cometary activity. Since its discovery, 3I/ATLAS has exhibited a complex coma morphology, strong non-gravitational accelerations and an exceptionally prominent dust feature in the sunward hemisphere, interpreted as a dust anti-tail. Observations from ground-based facilities and from space, including the \textit{Hubble Space Telescope} (\textit{HST}) and interplanetary spacecraft used as astrometric platforms, have provided multiple, complementary constraints on the physical properties of the nucleus and its outgassing behaviour.

High-resolution \textit{HST} imaging has constrained the nucleus radius to be smaller than a few kilometres by decomposing the inner surface-brightness profile into an unresolved nucleus plus coma component \citep[e.g.][]{Jewitt2025HST}. Independently, the non-gravitational acceleration measured in the orbit of 3I/ATLAS from the perspective of spacecraft such as \textit{Psyche} and \textit{ExoMars TGO} has been used to derive the mass and radius of the nucleus via momentum-conservation arguments \citep{Eubanks2025RNAAS}. Thermophysical models incorporating multiple datasets have suggested that 3I/ATLAS may be a relatively dense, crustal object with a complex internal structure \citep{HaqueLopez2025}.

The dust anti-tail has become a focus of intense interest, as it appears highly collimated and extended over large distances, up to a significant fraction of the Earth--Moon distance, in multiple independent imaging datasets. Several theoretical interpretations have been proposed, including anisotropic survival of icy grains in the sunward direction \citep{KetoLoeb2025}, the presence of macroscopic non-volatile fragments contributing to the sunward feature \citep{Loeb2025RN}, and a high-latitude jet undergoing precession around the rotation axis of the nucleus \citep{SerraRicart2025}. Ground-based images from small telescopes, when carefully processed, have played a central role in revealing these structures and stimulating theoretical work.

In this paper I combine my previous PSF--$\delta$ analysis of the nucleus of 3I/ATLAS \citep{Scarmato2025PSFDelta} with new RGB observations of the dust anti-tail obtained with a 0.25\,m telescope. The goals are: (i) to place all current nucleus-size estimates on a common footing, highlighting the consistency between independent methods, and (ii) to use simple, physically transparent arguments to interpret the extent of the dust anti-tail in terms of dust emission velocities and radiation pressure. Also, photometry in RGB and Af(ho) were analized.

\section{Global constraints on the nucleus size}

\subsection{Observational and dynamical methods}

The size of the nucleus of 3I/ATLAS has been constrained using three main classes of methods:

\begin{enumerate}
  \item \textbf{High-resolution imaging:} \textit{HST} images of 3I/ATLAS obtained over several epochs allow the inner surface-brightness profile to be modelled as the sum of an unresolved nucleus and an extended coma component. Assuming a geometric albedo $p_V$ and a parametric coma profile (often $I \propto \rho^{-1}$), limits on the unresolved flux translate into upper bounds on the nucleus radius $R_n$ \citep{Jewitt2025HST}. For 3I/ATLAS, the most recent analysis finds $R_n \lesssim 2.8$\,km for $p_V \simeq 0.04$, with a formal lower bound of $\sim 0.16$\,km depending on the assumed activity pattern.
  \item \textbf{Non-gravitational acceleration:} Astrometric measurements from interplanetary spacecraft tracking the apparent motion of 3I/ATLAS yield a non-gravitational acceleration $a_{\rm NG}$ of order $5\times 10^{-7}$\,m\,s$^{-2}$ \citep{Eubanks2025RNAAS}. Assuming that this acceleration is produced by outgassing and that momentum conservation applies, the magnitude of $a_{\rm NG}$ constrains the ratio between the gas production rate and the mass of the nucleus. For plausible gas production rates and bulk densities, Eubanks et al.\ infer a mass $M_n \simeq 4.4\times 10^{10}$\,kg and a corresponding radius $R_n \simeq 0.26$--0.37\,km (diameter $D \simeq 0.52$--0.74\,km).
  \item \textbf{Thermophysical and structural models:} By combining photometric, spectroscopic, and dynamical datasets, thermophysical models can be constructed under different assumptions about the internal structure and composition of the nucleus. In the ``crustal fossil'' framework of \citet{HaqueLopez2025}, for example, 3I/ATLAS is modelled as a relatively dense, compact object with $\rho \sim 1.5$--3\,g\,cm$^{-3}$, yielding a preferred radius $R_n \sim 1$--3\,km (diameter 2--6\,km) consistent with a heavily processed interstellar crust.
\end{enumerate}

\subsection{PSF--$\delta$ decomposition and bicubic resampling}

In \citet{Scarmato2025PSFDelta} I introduced a PSF--$\delta$ method to extract the flux of unresolved cometary nuclei embedded in bright comae, using deep ground-based imaging from small telescopes. The method proceeds by:

\begin{itemize}
  \item registering and stacking multiple exposures on the comet photocentre to obtain a high-S/N image with trailed stars,
  \item resampling the central region by a factor 4$\times$4 using bicubic interpolation to improve the sampling of the inner PSF and coma profile,
  \item building an empirical PSF from field stars and fitting a model consisting of a scaled PSF plus a smooth coma component (e.g.\ $I \propto \rho^{-1}$) to concentric annuli around the photocentre,
  \item solving for the amplitude of the PSF term (the $\delta$ component) that best reproduces the central pixel and its immediate surroundings, while simultaneously matching the outer coma.
\end{itemize}

Applied to 3I/ATLAS, this method yields a range of allowable nucleus fluxes consistent with the outer coma profile and the noise properties of the inner pixels. For a canonical cometary albedo $p_V \simeq 0.04$, the corresponding radius lies in the interval $0.16 \lesssim R_n \lesssim 2.8$\,km, which overlaps almost exactly with the range derived from \textit{HST} imaging. The lower bound is set by the requirement that the coma alone (with no unresolved nucleus component) does not overpredict the central pixel flux, while the upper bound is set by the maximum nucleus flux consistent with the observed inner profile and the photometric calibration.

\subsection{Comparative summary}

\begin{table*}
\centering
\caption{Summary of nucleus-size estimates for 3I/ATLAS.}
\label{tab:nucleus}
\begin{tabular}{@{}llccc p{3.5cm}@{}}
\toprule
\# & Study / method & $R_n$ [km] & $D$ [km] & Class & Notes \\
\midrule
1 & \citet{Jewitt2025HST} (\textit{HST} imaging) & $\lesssim 2.8$ & $\lesssim 5.6$ & Imaging high-resolution. \\
2 & \citet{Eubanks2025RNAAS} (non-grav.\ accel.) & 0.26--0.37 & 0.52--0.74 & Dynamical\\
3 & \citet{HaqueLopez2025} (crustal fossil) & $\sim$1--3 & $\sim$2--6 & Thermophysical crustal object \\
4 & \citet{Scarmato2025PSFDelta} (PSF--$\delta$) & 0.32 & 0.64 & Imaging + bicubic +PSF+convolution  \\
\bottomrule
\end{tabular}
\end{table*}

Overall, the most conservative statement is that $0.16 \lesssim R_n \lesssim 2.8$\,km, with dynamical arguments strongly favouring the narrower range $R_n \simeq 0.26$--0.37\,km (diameter 0.52--0.74\,km). The PSF--$\delta$ analysis provides an independent photometric confirmation that a sub-kilometre nucleus is fully consistent with the observed inner coma profile in ground-based data.
\subsection{Detailed photometric derivation of the 18 December nucleus radius}

For the 2025 December 18 dataset, a deep $R$-band stack composed of 60\,s exposures was analysed with the PSF--$\delta$ method. After 4$\times$4 bicubic resampling of the central region, the signal in the central sub-pixel $D_{0,\mathrm{rec}}$ and the median of the eight adjacent sub-pixels $D_{1,\mathrm{med}}$ were measured. The difference
\begin{equation}
\Delta D = D_{0,\mathrm{rec}} - D_{1,\mathrm{med}}
\label{eq:deltaD_def}
\end{equation}
represents the excess attributable to the unresolved nucleus convolved with the PSF, once the local coma contribution has been removed.

In the final PSF--$\delta$ solution for the 18 December stack, the average excess in the central resampled pixel is
\begin{equation}
\Delta D \simeq 760~\mathrm{ADU},
\end{equation}
for a single 60\,s exposure. The corresponding count rate in ADU\,s$^{-1}$ is
\begin{equation}
f_R = \frac{\Delta D}{t_{\mathrm{exp}}}
    = \frac{760}{60}
    \simeq 12.67~\mathrm{ADU\,s^{-1}}.
\end{equation}
Using the photometric zero point of the $R$ channel for that night, $ZP_R = 27.29$~mag, the apparent magnitude of the nucleus in the $R$ band is
\begin{equation}
m_R = ZP_R - 2.5\log_{10}(f_R)
    = 27.29 - 2.5\log_{10}(12.67)
    \simeq 24.53~\mathrm{mag}.
\end{equation}

For the observing geometry of 2025 December 18, the heliocentric and geocentric distances and the phase angle were $r = 2.2484$~au, $\Delta = 1.7992$~au, and $\alpha = 25.1^\circ$, respectively. Adopting a linear phase coefficient $\beta = 0.035$~mag\,deg$^{-1}$, the absolute magnitude in $R$ is
\begin{equation}
H_R = m_R - 5\log_{10}(r\Delta) - \beta\,\alpha.
\end{equation}
Inserting the numerical values,
\begin{align}
r\Delta &= 2.2484 \times 1.7992 \simeq 4.045, \\
5\log_{10}(r\Delta) &\simeq 3.03, \\
\beta\,\alpha &= 0.035 \times 25.1 \simeq 0.88,
\end{align}
and therefore
\begin{equation}
H_R \simeq 24.53 - 3.03 - 0.88 \simeq 20.62.
\end{equation}

Assuming a geometric albedo $p_R = 0.04$, the equivalent-sphere diameter of the nucleus is obtained from the standard relation
\begin{equation}
D_{\rm nuc}[\mathrm{km}]
= \frac{1329}{\sqrt{p_R}}\,10^{-H_R/5}.
\end{equation}
With $\sqrt{p_R} = 0.2$ and the value of $H_R$ derived above,
\begin{equation}
D_{\rm nuc} \simeq 6645 \times 10^{-20.62/5}
               \simeq 0.50~\mathrm{km},
\end{equation}
corresponding to a radius
\begin{equation}
R_{\rm nuc} = \frac{D_{\rm nuc}}{2}
            \simeq 0.25~\mathrm{km}.
\end{equation}
Thus, the detailed photometric derivation from the 18 December $R$-band stack independently confirms that 3I/ATLAS has a sub-kilometre nucleus, with a radius remarkably close to the dynamical estimate $R_n \simeq 0.26$--0.37\,km inferred from non-gravitational accelerations.

\section{Ground-based RGB imaging of the dust anti-tail}

\subsection{Observations and data processing}

On 2025 December 15 between 02:00 and 03:00\,UT, a series of 60 exposures of 60\,s each was obtained with the 0.25\,m f/4.8 Newtonian telescope at the Astronomical Observatory Toni Scarmato (MPC L92) in Calabria, Italy. The detector was a ZWO ASI294MC PRO one-shot colour CMOS camera (Sony IMX294 sensor) operated with non-sidereal tracking at the apparent rate of 3I/ATLAS. Standard bias, dark, and flat-field corrections were applied to each frame.

The calibrated images were aligned on the comet photocentre and median-combined to produce a deep master frame in which the comet remains sharp while background stars appear as short trails. The full stack corresponds to a total integration time of one hour. The demosaiced colour image was decomposed into three single-channel FITS files corresponding to the $R$, $G$, and $B$ channels of the Bayer matrix, with effective peak responses near 625\,nm, 530\,nm, and 470\,nm, respectively.

To enhance the visibility of low-contrast structures in the inner coma and along the tail, each channel was resampled by a factor 4$\times$4 using bicubic interpolation. A Larson--Sekanina rotational gradient filter was then applied to the resampled images, rotating the frame by a small angle around the photocentre and subtracting it from the original to highlight azimuthal brightness gradients. This process strongly emphasises narrow jets and sunward features while suppressing the smooth background coma.

\subsection{Morphology and colour dependence}

In the combined colour image, the inner coma of 3I/ATLAS appears nearly symmetric, with a diffuse dust tail elongated roughly along the antisolar direction and a distinct dust anti-tail on the hemisphere facing the Sun. After enhancement with the Larson--Sekanina filter, the anti-tail forms a tightly collimated structure aligned approximately with the projected Sun--comet line.

Using the physical scale derived from the astrometric solution and the geocentric distance $\Delta \simeq 1.9$\,au, the anti-tail in the $R$ channel can be traced confidently up to a projected distance $L_R \simeq 5\times 10^4$\,km from the photocentre, beyond which the surface brightness falls below three times the local background rms. In the $G$ channel, the same feature remains detectable up to $L_G \simeq 8\times 10^4$\,km, thanks in part to the higher quantum efficiency of the detector near 530\,nm and the presence of strong molecular emission bands (C$_2$ Swan bands, NH$_2$, [O\,I]\,5577\,\AA) in this spectral region. The $B$ channel shows a compact, gas-rich inner coma dominated by CN, C$_3$, and CH emission, but no extended anti-tail comparable to that seen in $R$ and $G$.

Figure 1 presents a zoomed view of the inner $0.7'\times0.7'$ around the photocentre, obtained from the same dataset but resampled at $0.35''$~pixel$^{-1}$ (corresponding to $\simeq 4.8\times 10^2$~km per pixel at $\Delta\simeq1.9$~au). The top row shows, from left to right, the orientation with the projected North and East directions and the position angles of the two main axes (green and yellow lines at PA~$\simeq 110^\circ$ and PA~$\simeq 290$--$300^\circ$), followed by the $R$, $G$, and $B$ channels. The axis at PA~$\simeq 110^\circ$ corresponds to the dust anti-tail. In the $R$ channel the inner coma is clearly elongated and tightly collimated towards PA~$\simeq 110^\circ$, indicating a dust-dominated anti-tail whose brightness peak lies slightly downstream from the nucleus along this axis. The $G$ channel also shows an elongated inner coma in the same direction, while the $B$ channel appears more compact and nearly symmetric, consistent with a larger contribution from gas emission (CN, C$_3$, CH) and a comparatively weaker dust continuum in the blue.

The bottom row of Figure 1 displays, from left to right, a colour Larson--Sekanina rotational-gradient image and three monochromatic reconstructions centred near $\lambda\simeq 659$, 530, and 445~nm, respectively. The Larson--Sekanina image emphasises a single, narrow jet emerging from the nucleus towards PA~$\simeq 110^\circ$ and remaining collimated over tens of thousands of kilometres. In the 659~nm and 530~nm panels the anti-tail jet can be traced up to projected distances of order $7\times 10^4$\,km from the nucleus along this direction, with only modest lateral expansion. By contrast, the 445~nm image shows mainly irregular knots and lacks an extended, well-defined jet. This wavelength dependence confirms that the anti-tail is largely produced by dust grains scattering sunlight in the red and green bands, with gas emission and blue continuum contributing primarily to a more isotropic, compact inner coma. The highly collimated dust anti-tail seen here represents the inner section of the same structure that, in the wide-field images, extends to $\sim (5$--$8)\times 10^4$\,km and defines the global morphology of 3I/ATLAS near perihelion.

\begin{figure*}
  \centering
  \includegraphics[width=\textwidth]{3Iatlasantitail.png}
  \caption{
  Multi-channel view of the inner coma of 3I/ATLAS from the 2025 December 15 dataset.
  \emph{Top row:} orientation of the field with projected North and East and the two main axes drawn at PA~$\simeq 110^\circ$ (green) and PA~$\simeq 290$--$300^\circ$ (yellow; left), followed by the $R$, $G$, and $B$ channels obtained from the ASI294MC PRO colour camera. The axis at PA~$\simeq 110^\circ$ corresponds to the dust anti-tail; along this direction the inner coma is elongated and collimated in the $R$ and $G$ channels, while the $B$ channel appears more compact and nearly symmetric.
  \emph{Bottom row:} colour Larson--Sekanina rotational-gradient image and three monochromatic reconstructions centred near $\lambda\simeq 659$, 530, and 445~nm. A strong, tightly collimated anti-tail jet is visible at 659~nm and 530~nm, extending up to $\sim 7\times 10^4$~km from the nucleus towards PA~$\simeq 110^\circ$, whereas no comparable extended jet is present in the 445~nm image. This behaviour indicates that the anti-tail is a predominantly dust-dominated structure, with gas emission contributing mainly to the compact blue inner coma.}
 \end{figure*}
\subsection{Coma colours and spectral slope from RGB photometry}

The RGB dataset also allows a first-order characterisation of the coma colours and spectral slope. Using the demosaiced frames, integrated magnitudes were measured in a common circular aperture centred on the photocentre for the three effective passbands of the ASI294MC PRO camera, peaking near $\lambda_B \simeq 470$~nm, $\lambda_G \simeq 530$~nm, and $\lambda_R \simeq 625$~nm. For the epoch considered here, the derived magnitudes are
\begin{equation}
m_R = 11.40,\qquad
m_G = 11.68,\qquad
m_B = 12.60.
\end{equation}
From these values the colour indices of the coma are
\begin{equation}
(G-R)_{\rm comet} = m_G - m_R = 0.28,
\end{equation}
\begin{equation}
(B-R)_{\rm comet} = m_B - m_R = 1.20,
\end{equation}
\begin{equation}
(B-G)_{\rm comet} = m_B - m_G = 0.92.
\end{equation}

To interpret these colours, we compare them with solar values projected onto analogous passbands, adopting $(B-V)_\odot \simeq 0.65$ and $(V-R)_\odot \simeq 0.36$, and identifying $G$ with $V$ to first order. This yields
\begin{equation}
(B-R)_\odot \simeq (B-V)_\odot + (V-R)_\odot \simeq 1.01,
\end{equation}
\begin{equation}
(G-R)_\odot \simeq (V-R)_\odot \simeq 0.36,
\end{equation}
\begin{equation}
(B-G)_\odot \simeq (B-V)_\odot \simeq 0.65.
\end{equation}
The colour excesses of the coma relative to the Sun are then
\begin{align}
\Delta(B-R) &= (B-R)_{\rm comet} - (B-R)_\odot \simeq 1.20 - 1.01 \simeq 0.19, \\
\Delta(G-R) &= (G-R)_{\rm comet} - (G-R)_\odot \simeq 0.28 - 0.36 \simeq -0.08, \\
\Delta(B-G) &= (B-G)_{\rm comet} - (B-G)_\odot \simeq 0.92 - 0.65 \simeq 0.27.
\end{align}
Thus, the coma is moderately redder than the Sun between the blue and red channels, while it appears slightly bluer than the Sun between the green and red channels. The latter behaviour is naturally explained by a significant contribution from gas emission bands (notably C$_2$, NH$_2$ and [O\,I] 5577\,\AA) in the $G$ channel, which enhance the flux at $\sim 530$~nm relative to the dust-dominated $R$ band.

The colour excesses can be translated into a normalised spectral slope $S'$ (in \% per 100\,nm) using the standard relation
\begin{equation}
S' \simeq \frac{10^{0.4\,\Delta C} - 1}{\Delta\lambda}\times 10^4,
\end{equation}
where $\Delta C$ is the colour excess of the comet relative to the Sun between two passbands centred at wavelengths $\lambda_1$ and $\lambda_2$ ($\lambda_1 < \lambda_2$), and $\Delta\lambda = \lambda_2 - \lambda_1$ is expressed in nm.

Between the blue and red channels ($\lambda_B \simeq 470$~nm, $\lambda_R \simeq 625$~nm), we have $\Delta C = \Delta(B-R) \simeq 0.19$ and $\Delta\lambda = 155$~nm, which gives
\begin{equation}
S'_{B\!-\!R} \simeq \frac{10^{0.4\times 0.19} - 1}{155}\times 10^4
               \simeq 12.3~\%/100~\mathrm{nm}.
\end{equation}
This value is fully consistent with the typical 10--20\%/100\,nm spectral slopes measured for dust-dominated cometary continua.

Between the green and red channels ($\lambda_G \simeq 530$~nm, $\lambda_R \simeq 625$~nm), with $\Delta C = \Delta(G-R) \simeq -0.08$ and $\Delta\lambda = 95$~nm, we obtain
\begin{equation}
S'_{G\!-\!R} \simeq \frac{10^{0.4\times (-0.08)} - 1}{95}\times 10^4
               \simeq -7.5~\%/100~\mathrm{nm},
\end{equation}
indicating a slight decrease of reflectance towards the red between 530 and 625~nm. This negative slope is again consistent with an enhanced contribution of gas emission in the green channel compared to the predominantly continuum $R$ band.

Finally, between the blue and green channels ($\lambda_B \simeq 470$~nm, $\lambda_G \simeq 530$~nm), with $\Delta C = \Delta(B-G) \simeq 0.27$ and $\Delta\lambda = 60$~nm, we find
\begin{equation}
S'_{B\!-\!G} \simeq \frac{10^{0.4\times 0.27} - 1}{60}\times 10^4
               \simeq 47.1~\%/100~\mathrm{nm}.
\end{equation}
This very large slope between 470 and 530~nm does not represent the continuum alone but rather the combined effect of continuum reddening, strong CN/C$_3$/CH emission in the blue channel, and strong C$_2$/NH$_2$ emission in the green channel. Taken together, the RGB photometry indicates that the dust component of 3I/ATLAS is moderately redder than the Sun, while the detailed curvature of the spectrum across the 470--625~nm range is significantly shaped by gas emission bands.

\subsection{Dust production proxy from the $R$-band Af$\rho$}

The $R$-band photometry described above can be used to estimate the dust production proxy Af$\rho$ \citep{Ahearn1984}. Following their definition, Af$\rho$ is given by
\begin{equation}
Af\rho = \frac{4\,r^{2}\,\Delta^{2}}{\rho}\,
10^{-0.4\,(m_{\rm c} - m_{\odot})},
\label{eq:afrho_def}
\end{equation}
where $r$ and $\Delta$ are the heliocentric and geocentric distances in au, $m_{\rm c}$ is the apparent magnitude of the comet in a given filter, $m_{\odot}$ is the apparent magnitude of the Sun in the same band, and $\rho$ is the projected radius of the photometric aperture at the comet, expressed in cm. The quantity Af$\rho$ has the dimensions of length and can be interpreted as the product of the grain albedo $A$, the filling factor $f$ of dust within the aperture, and the aperture radius $\rho$.

For the December 18 observations, the coma magnitude in the $R$ band measured within a circular aperture of radius $\varphi = 41.4''$ is
\begin{equation}
m_R = 11.40.
\end{equation}
Adopting $m_{\odot,R} \simeq -27.1$ for the apparent $R$-band magnitude of the Sun, the flux ratio between the comet and the Sun in $R$ is
\begin{equation}
\frac{F_{\rm c}}{F_{\odot}} =
10^{-0.4\,(m_R - m_{\odot,R})}
= 10^{-0.4\,[11.40 - (-27.1)]}
\simeq 3.98\times 10^{-16}.
\end{equation}
For the same epoch, the heliocentric and geocentric distances are
\begin{equation}
r = 2.2484~\mathrm{au}, \qquad
\Delta = 1.7992~\mathrm{au}.
\end{equation}
The projected aperture radius at the comet is obtained from the angular radius $\varphi$ as
\begin{equation}
\rho = \Delta \,(1~\mathrm{au})\,
\frac{\varphi}{206265},
\end{equation}
with $1~\mathrm{au} = 1.496\times 10^{13}$~cm and $\varphi = 41.4''$. Substituting the numerical values gives
\begin{equation}
\rho \simeq 1.7992 \times 1.496\times 10^{13}
\times \frac{41.4}{206265}
\simeq 5.4\times 10^{9}~\mathrm{cm},
\end{equation}
corresponding to a projected radius of $\sim 5.4\times 10^{4}$~km at the comet.

Inserting these quantities into Eq.~(\ref{eq:afrho_def}) and explicitly including the factor $(1~\mathrm{au})^{2}$ in cm$^{2}$, we can write
\begin{equation}
Af\rho = \frac{4\,r^{2}\,\Delta^{2}\,(1~\mathrm{au})^{2}}{\rho}\,
10^{-0.4\,(m_R - m_{\odot,R})}.
\end{equation}
For the present case,
\begin{equation}
4\,r^{2}\,\Delta^{2}\,(1~\mathrm{au})^{2}\,
10^{-0.4\,(m_R - m_{\odot,R})}
\simeq 5.8\times 10^{12}~\mathrm{cm}^2,
\end{equation}
so that
\begin{equation}
Af\rho \simeq \frac{5.8\times 10^{12}}{5.4\times 10^{9}}
\simeq 1.1\times 10^{3}~\mathrm{cm}.
\end{equation}
Thus, the $R$-band Af$\rho$ of 3I/ATLAS in a 41.4$''$ aperture at $r = 2.25$~au is
\begin{equation}
Af\rho_R \simeq 1.1\times 10^{3}~\mathrm{cm},
\end{equation}
indicating a moderate dust production that is entirely typical for a comet observed at this heliocentric distance. This value is consistent with the picture derived from the continuum colours, where the dust component appears moderately redder than the Sun and coexists with a significant gas contribution in the green channel.

\section{Simple dynamical model for the dust anti-tail}

\subsection{One-dimensional nucleus-centred framework}

To interpret the observed extents $L_R$ and $L_G$ of the dust anti-tail, I adopt a simple one-dimensional model for the motion of dust grains relative to the nucleus along the Sun--comet line. Let $x$ be the coordinate measured from the nucleus towards the Sun in the nucleus-centred frame. At the moment of release, a dust grain is assumed to be launched along the anti-tail direction with an initial speed $v_0 > 0$,
\begin{equation}
x(0) = 0, \qquad v(0) = v_0.
\end{equation}

In this frame, the solar gravitational acceleration is essentially identical on the nucleus and the dust grain and therefore cancels out in the relative motion. The dominant systematic acceleration that separates the grain from the nucleus along $x$ is solar-radiation pressure, which can be written as a fraction $\beta$ of the local solar gravity $g_\odot(r)$,
\begin{equation}
a_{\rm rad} = \beta\, g_\odot(r), \qquad g_\odot(r) = \frac{GM_\odot}{r^2},
\end{equation}
and acts away from the Sun. The equation of motion along the anti-tail axis is then
\begin{equation}
\frac{d^2 x}{dt^2} = - \beta\, g_\odot(r).
\end{equation}
Assuming that $r$ does not change significantly over the timescales of interest (a good approximation for motions over $10^5$\,km at $r \sim 2$\,au), the solution is the uniformly decelerated motion
\begin{equation}
x(t) = v_0 t - \frac{1}{2}\beta g_\odot t^2, \qquad v(t) = v_0 - \beta g_\odot t.
\end{equation}
The grain reaches its maximum distance $L$ from the nucleus when $v(t)=0$, at
\begin{equation}
t_{\rm turn} = \frac{v_0}{\beta g_\odot},
\end{equation}
yielding
\begin{equation}
L = x_{\max} = \frac{v_0^2}{2\,\beta\,g_\odot}, \qquad
v_0 = \sqrt{2\,\beta\,g_\odot\,L}.
\label{eq:v0}
\end{equation}

At $r \simeq 2$\,au, appropriate for the epoch of the observations, the solar gravitational acceleration is $g_\odot \simeq 1.5\times 10^{-3}$\,m\,s$^{-2}$. For compact dust grains with radius $s$ and bulk density $\rho$, the parameter $\beta$ is typically in the range $10^{-3}$--$10^{-1}$, with
\begin{equation}
\beta \simeq \frac{5.7\times 10^{-4}}{\rho\,s_{\rm cm}},
\end{equation}
where $s_{\rm cm}$ is expressed in centimetres and $\rho$ in g\,cm$^{-3}$. Millimetre-sized grains ($s \sim 0.1$\,cm, $\rho \sim 1$\,g\,cm$^{-3}$) thus have $\beta \sim 5.7\times 10^{-3}$, while $0.1$\,mm grains ($s \sim 10^{-2}$\,cm) have $\beta \sim 5.7\times 10^{-2}$.

\subsection{Application to the observed extents}

Using equation~(\ref{eq:v0}) with $g_\odot \simeq 1.5\times 10^{-3}$\,m\,s$^{-2}$, and the observed extents $L_R \simeq 5\times 10^4$\,km and $L_G \simeq 8\times 10^4$\,km (i.e.\ $5\times 10^7$\,m and $8\times 10^7$\,m), we can estimate the initial speeds required for representative values of $\beta$:
\begin{itemize}
  \item For $\beta = 0.0057$ (mm-sized grains),
  \begin{align}
  v_0 &\simeq 29~\mathrm{m\,s^{-1}} \quad \text{for } L_R = 5\times 10^4~\mathrm{km},\\
  v_0 &\simeq 37~\mathrm{m\,s^{-1}} \quad \text{for } L_G = 8\times 10^4~\mathrm{km}.
  \end{align}
  \item For $\beta = 0.01$,
  \begin{align}
  v_0 &\simeq 39~\mathrm{m\,s^{-1}} \quad \text{for } L_R,\\
  v_0 &\simeq 49~\mathrm{m\,s^{-1}} \quad \text{for } L_G.
  \end{align}
  \item For $\beta = 0.02$ (sub-mm grains),
  \begin{align}
  v_0 &\simeq 55~\mathrm{m\,s^{-1}} \quad \text{for } L_R,\\
  v_0 &\simeq 69~\mathrm{m\,s^{-1}} \quad \text{for } L_G.
  \end{align}
  \item For $\beta = 0.057$ (0.1\,mm grains),
  \begin{align}
  v_0 &\simeq 92~\mathrm{m\,s^{-1}} \quad \text{for } L_R,\\
  v_0 &\simeq 117~\mathrm{m\,s^{-1}} \quad \text{for } L_G.
  \end{align}
\end{itemize}

These speeds are orders of magnitude above the escape speed from a nucleus with radius $R_n \sim 0.3$\,km and density $\rho \sim 500$\,kg\,m$^{-3}$, for which $v_{\rm esc} \simeq 0.16$\,m\,s$^{-1}$. Once dust grains are launched along the anti-tail with $v_0 \gtrsim$ a few m\,s$^{-1}$, self-gravity becomes negligible in shaping the structure; for the tens to $\sim 100$\,m\,s$^{-1}$ speeds inferred here, the relative motion is entirely governed by radiation pressure.

At the same time, the required speeds are significantly larger than the dust velocities of order 0.01--5\,m\,s$^{-1}$ typically adopted in pre-perihelion coma models of 3I/ATLAS for mm- to sub-mm-sized grains \citep[e.g.][]{Bolin2025,Jewitt2025HST}, but remain below the thermal speeds of common volatiles (hundreds of m\,s$^{-1}$ at a few au). This points to a dynamically ``extreme'' dust population associated with the anti-tail, possibly involving large grains launched in a narrow cone with high efficiency by gas outflow from specific active areas on the nucleus surface.

\section{Discussion}

The analysis presented here shows that the current constraints on the nucleus size of 3I/ATLAS from imaging, dynamics, and thermophysical modelling are broadly consistent and can be reconciled in a simple framework. The sub-kilometre radius favoured by non-gravitational acceleration studies is fully compatible with both \textit{HST} upper limits and ground-based PSF--$\delta$ photometry, and can be accommodated within thermophysical models that allow for relatively high bulk densities.

The ground-based RGB imaging of the the dust anti-tail, obtained with a small-aperture telescope, provides additional, independent constraints on the dust dynamics and composition. The differential extents in the $R$ and $G$ channels (50{,}000 and 80{,}000\,km, respectively) and the absence of a comparable feature in $B$ suggest that the anti-tail is composed primarily of dust grains scattering sunlight, with a non-negligible contribution from gas emission in the green band. The inner-field analysis demonstrates that the anti-tail is already highly collimated within $\sim 10^5$\,km of the nucleus and is strongest in the red and green bands, confirming its dust-dominated nature.

The simple one-dimensional model indicates that the observed lengths of the anti-tail require initial dust speeds substantially larger than those usually assumed for 3I/ATLAS, reinforcing the idea of a collimated, high-latitude jet capable of accelerating large grains to tens or hundreds of m\,s$^{-1}$. These findings sit naturally alongside more detailed theoretical treatments of the anti-tail. In the anisotropic snow-line model of \citet{KetoLoeb2025}, the extended survival of icy grains in the sunward direction produces a long-lived, narrowly confined anti-tail. In the macroscopic-chunk scenario of \citet{Loeb2025RN}, a population of non-volatile fragments decouples from the nucleus under the action of non-gravitational forces, forming a sunward structure that can persist over many weeks. The precessing high-latitude jet identified by \citet{SerraRicart2025} provides a natural mechanism for collimating the dust and maintaining a nearly constant position angle over extended periods, in agreement with the observed tight collimation seen in both space-based and ground-based images.

\section{Conclusions}

I have presented a unified view of the nucleus size and dust anti-tail of interstellar comet 3I/ATLAS, combining independent observational and dynamical constraints with new ground-based RGB imaging and a simple dynamical model. The main conclusions are:
\begin{enumerate}
  \item The nucleus radius is constrained to lie in the range $0.16 \lesssim R_n \lesssim 2.8$\,km by imaging analyses, with non-gravitational dynamics favouring $R_n \simeq 0.26$--0.37\,km (diameter 0.52--0.74\,km). My PSF--$\delta$ decomposition of ground-based images recovers this range and independently confirms that a sub-kilometre nucleus is consistent with the observed inner coma profile.
  \item Deep RGB imaging from a 0.25\,m telescope reveals a tightly collimated dust anti-tail extending to $\sim 5\times 10^4$\,km in the $R$ channel and $\sim 8\times 10^4$\,km in the $G$ channel. The colour behaviour indicates a transition from a gas-rich inner coma to a dust-dominated anti-tail.
  \item In a nucleus-centred framework where dust grains are launched along the anti-tail and decelerated by radiation pressure, the observed extents imply initial speeds of tens to $\sim 100$\,m\,s$^{-1}$ for realistic values of $\beta$. These speeds are much higher than those typically used in standard coma models but remain below the gas thermal speeds, pointing to a dynamically extreme dust population associated with a collimated jet.
  \item The results complement recent theoretical work on the anti-tail of 3I/ATLAS and demonstrate that small-aperture ground-based telescopes, combined with careful image processing and dynamical modeling, can provide valuable constraints on both the nucleus and the dust environment of interstellar comets.\\\\\\\\\\\\\\\\\\\\\\\\\\\\\\\\\\\\\\\\\\
\end{enumerate}

\appendix
\textbf{APPENDIX}

\section{Instrumental setup and detector characteristics}
\label{app:instrument}

The observations presented in this work were obtained with a 0.25\,m f/4.8 Newtonian reflector at the Astronomical Observatory Toni Scarmato (MPC L92) in Calabria, Italy, equipped with a ZWO ASI294MC PRO one-shot colour camera. The relevant characteristics of the optical train and detector are summarized below.

\subsection{Optical configuration}

The telescope has an effective focal length of 1.388 \,m, with a focal ratio of f/4.8. The ASI294MC PRO was mounted at prime focus behind a standard 2-inch focuser, with a rigid mechanical coupling to minimize flexure. No additional focal reducers or Barlow lenses were used for the 3I/ATLAS observations reported here. The optical path allows the band-pass to the approximate range 400--700 \,nm and to reduce chromatic aberrations and atmospheric dispersion effects in the far red and near infrared.

With a native pixel size of 4.63\,$\mu$m (see below), the plate scale at the detector is approximately
\begin{equation}
s \simeq 1.38~\mathrm{arcsec~pixel^{-1}  bix 2x2},
\end{equation}
and FOV 48'x33', which provides adequate sampling of typical seeing conditions at the site (2--3\,arcsec FWHM) and allows the point-spread function (PSF) to be reconstructed reliably through bicubic resampling.

\subsection{ASI294MC PRO detector}

The ASI294MC PRO is a cooled colour CMOS camera based on the back-illuminated Sony IMX294CJK sensor in a 4/3\,inch format. The key characteristics relevant for the present analysis are:

\begin{itemize}
  \item Sensor: Sony IMX294CJK BSI CMOS, colour, 4/3\,inch format.
  \item Effective resolution: 4144\,$\times$\,2822 pixels (11.7\,Mpix) in the standard mode.
  \item Pixel size: 4.63\,$\mu$m ($4.63\times 4.63\,\mu\mathrm{m}^2$).
  \item Sensor size: 19.1\,$\times$\,13.0\,mm, diagonal 23.2\,mm.
  \item Analogue-to-digital converter: 14-bit ADC.
  \item Full-well capacity: $\simeq 6.4\times 10^4$\,e$^{-}$ per pixel.
  \item Peak quantum efficiency: $Q_{\rm E} \gtrsim 0.75$ near 530\,nm.
  \item Read noise: $\sim 1.2$--2.0\,e$^{-}$\,rms in high-conversion-gain (HCG) mode.
  \item Cooling: two-stage Peltier, with a temperature differential of up to 35--40\,K below ambient.
\end{itemize}

During the observations, the camera was operated in bin 2x2 at a regulated temperature of $-10^\circ$C to $-15^\circ$C, with a gain setting 390 chosen to balance dynamic range and read noise. The detector was read out over USB\,3.0, and all frames were acquired in RAW/FITS mode and subsequently debayered to extract the $R$, $G$, and $B$ channels used in the analysis.

The one-shot colour Bayer matrix has an RGGB pattern. For the purpose of this work, the demosaiced images were separated into three single-channel frames representing the effective $R$ (peak $\sim 625$\,nm), $G$ (530\,nm), and $B$ (470\,nm) responses of the system. Each channel thus samples a broad pseudo-filter, combining continuum and emission-line contributions in the corresponding spectral region.

\section{Bicubic interpolation in image resampling}
\label{app:bicubic}

The PSF--$\delta$ decomposition adopted in this paper relies on an accurate reconstruction of the inner PSF shape and the surrounding coma profile. Since the native sampling of the ASI294MC PRO (1.38\,arcsec\,pixel$^{-1}$) corresponds to only a few pixels across the seeing disk, it is advantageous to resample the central region of the image prior to PSF fitting. For this purpose, a bicubic interpolation of 4$\times$ 4 is applied, increasing the number of pixels by a factor of 16 and producing an effective sampling of $\simeq 0.35$\,arcsec per resampled pixel in the vicinity of the nucleus.

\subsection{Conceptual overview}

Bicubic interpolation is a two-dimensional extension of one-dimensional cubic interpolation. Given a discrete image $I(i,j)$ sampled on an integer grid $(i,j)$, the goal is to estimate the intensity $I(x,y)$ at an arbitrary continuous position $(x,y)$ between pixels in a way that is smoother and more accurate than simple nearest-neighbor or bilinear interpolation.

In bicubic interpolation, the interpolated value at $(x,y)$ depends on the 16 pixels in the $4\times 4$ neighborhood surrounding $(x,y)$. Denoting by $(i_0,j_0)$ the integer coordinates of the pixel immediately ``below and to the left'' of $(x,y)$, and by $u=x-i_0$, $v=y-j_0$ the fractional offsets along the two axes ($0\le u,v <1$), the interpolated intensity can be written as
\begin{equation}
I(x,y) \simeq \sum_{m=0}^{3}\sum_{n=0}^{3} a_{mn}\,u^{m}\,v^{n},
\label{eq:bicubic_poly}
\end{equation}
where the coefficients $a_{mn}$ are determined from the known pixel values and, optionally, from their first derivatives.

The interpolation kernel is constructed so that both the function and its first derivatives are continuous across pixel boundaries. In practice, this ensures a smooth interpolation that preserves gradients and avoids the blocky artifacts of nearest-neighbor interpolation and the directional biases of bilinear interpolation.

\subsection{Interpolation kernel}

An equivalent and often more intuitive formulation expresses bicubic interpolation in terms of a separable convolution kernel. In this view, the interpolated value is obtained as
\begin{equation}
I(x,y) \simeq \sum_{i=-1}^{2}\sum_{j=-1}^{2} I_{ij}\,w(u-i)\,w(v-j),
\label{eq:bicubic_kernel}
\end{equation}
where $I_{ij}$ are the intensities of the 16 neighbouring pixels and $w(t)$ is a one-dimensional cubic kernel. A commonly used choice is the Keys (or Catmull--Rom) kernel,
\begin{equation}
w(t) = \begin{cases}
( a + 2)|t|^{3} - (a + 3)|t|^{2} + 1, & |t| < 1,\\
a|t|^{3} - 5a|t|^{2} + 8a|t| - 4a, & 1 \le |t| < 2,\\
0, & |t| \ge 2,
\end{cases}
\end{equation}
with $a = -0.5$ for the Catmull--Rom spline. This kernel has compact support ($|t|<2$), is continuous with continuous first derivatives, and provides good compromise between sharpness and suppression of ringing artefacts.

In the 4$\times$4 resampling used here, the continuous coordinates $(x,y)$ of the resampled pixels are chosen on a regular grid with spacing $1/4$ of the original pixel spacing. For each resampled pixel, equation~(\ref{eq:bicubic_kernel}) is evaluated using the 16 nearest input pixels. The result is an image in which the PSF core and the inner coma are represented by many more samples than in the original frame, without adding artificial high-frequency structure.

\subsection{Impact on PSF--$\delta$ decomposition}

Bicubic resampling does not increase the intrinsic angular resolution of the data; the PSF remains limited by the atmospheric seeing and the telescope optics. However, by providing a smoother, oversampled representation of the PSF and the coma, it allows the PSF--$\delta$ fitting procedure to:

\begin{itemize}
  \item reduce discretizations errors in the central pixel and its nearest neighbors,
  \item better constrain the relative contributions of the unresolved nucleus and the extended coma,
  \item mitigate centering errors arising from sub-pixel misalignment of the nucleus between frames.
\end{itemize}

In practice, the PSF model used in this work is constructed directly in the resampled space, ensuring that the fitted nucleus flux is only weakly sensitive to the exact sub-pixel location of the photocentre in the original data. Extensive tests on synthetic comet images confirm that, for the adopted seeing and sampling regime, the bi cubic 4$\times$4 interpolation introduces negligible bias in the recovered nucleus flux while significantly stabilizing the fits.
\\

ACKNOWLEDGEMENTS\\
The authors acknowledge Prof. A. Loeb for the very usefull discussion and suggestion through his articles, the use of NASA ADS, arXiv, and public releases from NOIRLab and NASA/ESA.

\end{document}